\documentclass[10pt, prd, amssymb,aps,twocolumn]{revtex4}

\begin{document}
\title{The Efficiency of Defect Production in Planar Superconductors and Liquid Crystals}
\author{R.\ J.\ Rivers}%
\email[E-mail:] {r.rivers@imperial.ac.uk}
\author{A. Swarup}%
\email[Email:]{amarendra.swarup@imperial.ac.uk}%
\address{Imperial College, London SW7 2AZ}
\date{\today }

\begin{abstract}
A recent experiment that sees the spontaneous creation of magnetic
flux on quenching high-$T_c$ films has shown that earlier null
results were a consequence of the lack of saturation of the
Zurek-Kibble causal bounds against which flux density was
measured. In this letter we estimate how efficient the production
of topological charge is in planar systems, both for the
aforementioned experiment (when flux measures topological charge)
and for an earlier experiment on planar liquid crystals. Agreement
is good.

\end{abstract}
\maketitle
\medskip

There have now been several experimental tests of the Zurek-Kibble
(ZK) causal bounds for the production of defects (e.g. vortices)
at continuous phase transitions. The ZK prediction is that the
separation of defects $\xi_{def}$, at the time of their formation,
satisfies a scaling law \cite {zurek1,zurek2,kibble1}
\begin{equation}
\xi _{def} = \frac{\xi_Z}{f}\equiv\frac{\xi _{0}}{f}\bigg(\frac{t _{Q}}{t_{0}%
}\bigg)^{\sigma },  \label{xidef}
\end{equation}
where $t_Q^{-1}$ is the rate at which the temperature is quenched
through its critical value $T_c$, $t_0$ is the relaxation time of
the long wavelength modes, and $\xi_0$ is the correlation length
at zero temperature. $\xi_{def}$ is measured in units of  $\xi_Z$,
the Zurek causal length, defined through (\ref{xidef}). $\xi_Z$ is
the minimum size of correlated domains after the implementation of
the transition\cite{zurek1,zurek2}. The efficiency coefficient
$f<1$ that determines the extent to which the causal bound is
saturated will depend on the system, but is generally assumed to
be $O(1)$. Typically, for mean-field condensed matter systems
describable by Ginzburg-Landau (GL) theory, the exponent $\sigma =
1/4$. Experiments in annular Josephson junctions\cite{roberto} and
non-linear optical systems \cite{florence} have confirmed this
value, where it is appropriate, within errors.

In this letter we begin by examining recent experiments to
detect Abrikosov-like vortices at a temperature quench in planar
high-$T_c$ superconductors by measuring the magnetic flux that is
spontaneously produced. The first of these experiments
\cite{technion} was null, with no flux observed above the noise in
the system. However, on increasing the quench rate by several
orders of magnitude spontaneous flux was produced
\cite{technion2}, but with low efficiency.

The experiments of Refs. \cite{technion} and \cite{technion2}
measure {\it net} magnetic flux i.e. net topological charge
$\Delta n$, where $\Delta n$ is the number of vortices $n_+$ {\it
minus} the number of antivortices $n_-$. As it stands, however,
Eq.\ref{xidef} only enables us to determine the total number $n =
n_+ + n_-$ of vortices {\it plus} antivortices. To make
predictions for net flux requires an understanding of
vortex-antivortex correlations. In the absence of explicit
knowledge about correlations, both Kibble \cite{kibble1,kibble2}
and Zurek \cite{zurek2} addressed this problem for $U(1)$
theories, with complex order parameter field $\phi = \rho
e^{i\alpha}$, by assuming that the phase angle $\alpha$ makes a
random walk around the perimeter of the region through which the
flux is measured, with step length $O(\xi_Z)$. This is
complemented by the 'geodesic rule', that the shortest paths are
assumed to be taken in field space so as to minimise gradient
energies. With these assumptions, the variance $\Delta N$ in the
net flux through a surface with perimeter $C$ is predicted to be
\begin{equation}
\Delta N = {\tilde f}\sqrt{\frac{C}{\xi_Z}} = {\tilde
f}\sqrt{\frac{C}{\xi_{0}}}\bigg(\frac{t _{Q}}{t_{0}%
}\bigg)^{-\sigma /2} \label{DeltaN}
\end{equation}
for $\sigma$ of (\ref{xidef}) with an overall efficiency factor
${\tilde f}$, that again depends on the system. The relationship
(\ref{DeltaN}) was originally proposed by Zurek \cite{zurek2} for
superconductors and for spontaneous superflow in $^4He$.

Our first goal is to show how (\ref{DeltaN}) arises from simple
dynamics, and to estimate the efficiency factor ${\tilde f}$ for
superconductors. Our second goal, postponed until later, is to
apply the same ideas to experiments \cite{digal,Ray} that measure
$\xi_{def}$ and $\Delta N$ in liquid crystals.

Care is needed in making comparisons between the two types of
system. In particular, for flux measurement in superconductors,
$\Delta N$ of (\ref{DeltaN}) can only be part of the story. We
also expect to see flux due to the freezing in of the long
wavelength modes of the electromagnetic field at the transition
\cite{rajantie}. In contrast to (\ref{DeltaN}), which implies
strong vortex-antivortex correlation, this shows strong
vortex-vortex correlation. However, for low $T_c$ and strong
type-II behaviour the effects due directly to the magnetic field
are negligible, as confirmed in numerical simulations
\cite{antunes,calzetta}.  Although larger for high-$T_c$, its
contribution to the variance in the net flux is still expected to
be relatively small \cite{rajantie2} in the experiments of
{\cite{technion,technion2}, and will be ignored. Further, since
their contribution is positive for the case in point, they can
only lead to an effective increase in efficiency.

Whereas low-$T_c$ superconductors are well-described by a $U(1)$
theory, high-$T_c$ superconductors are relatively poorly
understood. Arguably, the effective order parameter of the
high-$T_c$ superconductor is also a complex field $\phi$. This is
seen in an idealised $SO(5)$ model of high-$T_c$ superconductors
with a five-component 'rotor' order parameter field \cite{zhang},
three of whose components are the N\'{e}el vector of
antiferromagnetism and the remaining two the complex field
coupling to electromagnetism. The $SO(5)$ symmetry is explicitly
broken, restored only at critical doping. Classically, the order
parameter 'rotor' has fixed length, forcing vortices to have
antiferromagnetic cores \cite{arovas}, i.e. we have a generalised
$X-Y$ model (or, in the parlance of QFT, a non-linear
$\sigma$-model). However, fluctuations soften the constraint and,
unless the system is very close to critical doping \cite{alama},
normal Abrikosov vortices with winding number $n=\pm 1$ are
produced from the breaking of the electromagnetic $U(1)$ for a
type-II system. The symmetry-breaking groundstates of the theory in this case are the $S^1$ of the residual 'Mexican hat'
potential.

Eq.\ref{DeltaN} is tested in \cite{technion2} for $\sigma = 1/4$.
With $\sigma/2\approx 1/8$ so small, there is no way that scaling
can be accurately confirmed. However, the flux that is observed is
commensurate with (\ref{DeltaN}), provided ${\tilde f}\approx
1/16$, a low efficiency. It was noted in Ref.\cite{technion2}
that, if the superconductor were to behave as an X-Y model, for
which $\sigma\approx 0.15$, rather than $0.25$, we would expect an
order of magnitude more vortices and antivortices. Since we are
hardly seeing enough for the $U(1)$ GL case, we do not consider
this possibility further.  If, on the other hand, there was no
correlation between vortices and antivortices, $(\Delta N)^2$
would be proportional to area $A\propto C^2$, to give $\Delta
N\propto C/\xi_{def}\propto (t_Q/t_0)^{-\sigma}$, in contrast to
(\ref{DeltaN}). Equivalently, with the total number of vortices
and antivortices $N\propto A/\xi_{def}^2\propto (C/\xi_{def})^2$
this would imply $\Delta N\propto N^{1/2}$. This was originally
proposed in \cite{technion}, but the new data is totally
incompatible with such a possibility.

To show how (\ref{DeltaN}) is justified and that a low efficiency
is reasonable, we need microscopic field dynamics which,
inevitably, not only encode causality but also discriminate
between systems in a way that causal bounds cannot. Our
understanding of vortex formation is that, once the transition
temperature has been crossed, field ordering is implemented by the
growth in the amplitudes of the unstable long wavelength modes.
The vortices embody the frustration of the order parameter field as it
adopts different groundstate values in adjacent domains.

In this regard, we assume that field fluctuations can be taken to
be Gaussian until the time ${\bar t}$, say, when the back-reaction
can be considered to stop field growth.  For strongly damped
systems like those considered here it can be taken to be
effectively instantaneous. Specifically, if the symmetry breaking
scale is $|\phi | = \eta$, the transition is implemented by the
time ${\bar t}$ for which $\langle |\phi |^2\rangle_{\bar
t}\approx\eta^2$. Because of this we have the consequences of
Gaussian fluctuations imprinted on a subsequently highly
non-linear system.

 We
identify vortices by the (line) zeroes of the field at this time.
Writing the complex order-parameter field $\phi$ as $\phi =
(\phi_1 + i\phi_2)/\sqrt{2}$, the {\it total density} of zeroes in
the (12) plane (intersections of vortices plus antivortices) is
\begin{equation}
\bar{\rho}({\bf x}) = \delta^{2}[\phi ({\bf
x})]|\epsilon_{jk}\partial_{j} \phi_{1}({\bf x})
\partial_{k}\phi_{2}({\bf x})|. \label{rhobar}
\end{equation}
where $\delta^{2}[\phi ({\bf x})] = \delta[\phi_{1} ({\bf x})]
\delta[\phi_{2} ({\bf x})]$ and $\epsilon_{jk}$ is the
antisymmetric tensor.  For independently Gaussian $\phi_i$, the
ensemble average total density, to be matched to (\ref{xidef}),
is\cite{halperin,maz}
\begin{equation}
d(t) = -\frac{1}{2\pi}\bigg(\frac{G''(0,t)}{G(0,t)}\bigg).
\label{density}
\end{equation}
In (\ref{density}) $G(r,t) = \langle\phi ({\bf x})\phi^* ({\bf
0})\rangle_t$ is the field correlation function at time $t$ at
separation $r=|{\bf x}|$ and the derivatives are with respect to
$r$. It should also be noted that $d({\bar t})=(\xi_{def})^{-2}$
from (\ref{xidef}).

As we are considering planar systems, for simplicity we shall only
look at the properties of field zeroes in the (12) plane produced
by a quench. What is relevant here, for high-$T_c$ experiments, is
not the {\it total} density (\ref{rhobar}), but rather the {\it
topological} density of field zeroes of $\phi ({\bf x})$, which
determines net magnetic flux.  This is
\begin{equation}
\rho ({\bf x}) = \delta^{2}[\phi ({\bf
x})]\epsilon_{jk}\partial_{j} \phi_{1}({\bf x})
\partial_{k}\phi_{2}({\bf x}). \label{rho3}
\end{equation}
in which the modulus has been dropped.

Consider a superconductor film of area $A$ and circumference
$C$, enclosing $n_+$ vortices and $n_-$ antivortices. The
topological charge $\Delta n = n_+ - n_-$ will be zero on average,
$\langle\rho\rangle = 0$. However, there will be a nonzero
variance
\begin{equation}
(\Delta N)^2 = \int_{{\bf x},{\bf y}\in A}d^2x\,d^2y\,\langle\rho
({\bf x})\rho ({\bf y})\rangle_{\bar t}\label{DN}
\end{equation}
at time $t={\bar t}$ when vortices form.

Given the density $\rho$ of (\ref{rho3}) and the results of
\cite{halperin,maz}, it has been shown by one of us (RR)
\cite{RKK} that, if $g(r,t) = G(r,t)/G(0,t)$, then
\begin{equation}
(\Delta N)^{2}=
\frac{2C}{4\pi^2}\int_{0}^{\infty}dr\frac{g'^{2}(r,t)}{1 -
g^{2}(r,t)}. \label{step}
\end{equation}
This equation is the pivotal equation of this paper. The linear
dependence on $C$ is purely a result of the formative Gaussian
fluctuations. Although it could be argued that the 'geodesic rule'
is reasonable, when vortices have classical profiles, it is not
obviously the case at the early times that are relevant here, when
vortices are appearing from a tangle of line zeroes. Nonetheless,
we have recovered the consequences of this rule without needing to
mention geodesics.

On the other hand, it could be argued that the use of
(\ref{density}) to count vortices is equally unreasonable, since
the creation of classical vortices requires the non-Gaussian
interaction between $\phi_1$ and $\phi_2$. There is an answer to
this for global $U(1)$ theory, and our neglect of magnetic field
fluctuations suggests that this may provide a partial answer here.
Essentially, at time ${\bar t}$ when most of the field energy is
in field gradients, the energy ${\cal E}$ associated with the
gradient energy of the {\it Gaussian} field $\phi$ in a surface of
area A takes the form
\begin{equation}
{\cal E} = \int d^2x\,\langle|\nabla\phi |^2 \rangle_{\bar
t}\approx 2\pi \eta^2\int d^2x\,\langle\bar{\rho}({\bf
x})\rangle_{\bar t}
\end{equation}
on using the result (\ref{density}). That is, ${\cal E}\approx
2\pi\eta^2 N$, where $N = \langle n\rangle$ is the ensemble
average of $n=n_++n_-$, the total number of vortices  and
antivortices respectively. A vortex is a tube of false vacuum
whose energy/length is $\epsilon\sim 2\pi\eta^2$, showing that,
qualitatively, the energy matches that of a set of classical
defects whose number is the number of field zeroes (of Gaussian
fields) at the time at which defects are formed. It follows from
the Gaussian approximation that, to a first approximation,
dimensionality is not important if we ignore the frozen
fluctuations of the magnetic field, as we shall. For these, dimensionality
is important, and is a determinant in demonstrating their smallness \cite{rajantie2}.

There are further qualifications. For example, zeroes occur on all
scales, and $G(0, {\bar t})$ is infinite without a cutoff.
Provisionally, we introduce coarsegraining by hand, modifying
$G(r,t)$ by damping short wavelengths $l= O(\xi_0)$ in $D=2$
spatial dimensions as
\begin{equation}
G(r,t) = \int d \! \! \! / ^2 k\, e^{i{\bf k}.{\bf
x}}P(k,t)\,e^{-k^{2}l^{2}}. \label{Gl}
\end{equation}
In terms of the moments of $P(k,t)\,e^{-k^{2}l^{2}}$,
\begin{equation}
G_{n}(t) = \int_{0}^{\infty}dk\,k^{2n
-1}\,P(k,t)\,e^{-k^{2}l^{2}},
\end{equation}
the density $d$ of (\ref{density}) is the second moment
$d=G_2/2\pi G_1$. The cutoff is irrelevant once the 'Bragg peak'
at wavenumber $k_0\approx\xi_{def}^{-1}$ in the power has passed
to long wavelength modes. This has been discussed in detail
elsewhere \cite{rayulti,RKK} by one of us (RR).

We note that $(\Delta N)^{2}$ requires more than the lowest
moments of $P(k,t)$ needed for the total density $d$. A tedious
calculation \cite{RK} shows that, approximately
 \begin{equation}
 \frac{(\Delta N)^2}{C}\propto \frac{3G_3}{20G_2}
 -\frac{G_2}{12G_1}.
 \label{moments}
 \end{equation}
Nonetheless, as long as $k P(k,{\bar t})$
 is strongly peaked at $k_0\sim\xi_{def}^{-1}$ then, with only one
length scale, we recover (\ref{DeltaN}). Eqs. (\ref{density}) and
(\ref{moments}) show that the efficiency factors $f$ and ${\tilde
f}$ of (\ref{xidef}) and (\ref{DeltaN}) respectively are not
related directly. A low efficiency in producing total flux density
does not require a low efficiency in producing net flux.

In order to get an idea of the size of the efficiency ${\tilde f}$
for producing net flux we need a model for $g(r,{\bar t})$. The
vanishing of $\langle\phi\partial_i\phi^*\rangle$ requires that
$g'(0,t) = 0$. Empirically, we choose
\begin{equation}
g(r,{\bar t})\approx \exp (-r^2/2{\bar \xi}^2) \label{f}
\end{equation}
for $r$ sufficiently small to dominate the integral, for some
${\bar\xi} = O(\xi_Z)$ to be determined. At times ${\bar t}\gg
t_0$ such a Gaussian form is appropriate for the diffusion of
field ordering (independent of dimension $D$) that we would expect
from a dissipative system. Inserting this in (\ref{step}) gives
\begin{equation}
(\Delta N)^{2}\approx
\frac{C}{4\pi^2{\bar\xi}}\Gamma\bigg(\frac{3}{2}\bigg)\zeta\bigg(\frac{3}{2}\bigg)\approx
0.059\frac{C}{{\bar\xi}}. \label{step2}
\end{equation}
If ${\bar\xi} = \xi_Z$ we find ${\tilde f}\approx 1/4$, the value
obtained \cite{rudaz} by a naive discretisation of the phase angle
into multiples of $\pi/2$ and the 'geodesic rule'.

 For diffusive behaviour we expect
${\bar\xi}^2\propto{\bar t}$, independent of dimension. In turn,
we expect ${\bar t}= O(t_Z)$, where $t_Z$ is the Zurek causal
time, the first time at which the causal horizon is large enough
to accommodate defects.  Putting this together, in units of
$\xi_0$ and $t_0$, we would expect
\begin{equation}
{\bar\xi}^2\sim \xi_0^2 t_Z/t_0 \label{barxi}
\end{equation}
with a coefficient of proportionality $O(1)$.

In mean-field $t_Z\approx\sqrt{t_Qt_0}$ \cite{zurek2}. Inserting
this in (\ref{barxi}) and then, in turn, into (\ref{step2}),
recreates (\ref{DeltaN}) with $\sigma=1/4$, justifying the random
walk assumption of ZK in units of $\xi_Z$. This is our first
general result. As we noted earlier, Eq.\ref{DeltaN} is tested in
\cite{technion2} for $\sigma = 1/4$.

To fix the efficiency coefficient ${\tilde f}$ in ($\ref{DeltaN}$)
requires the coefficient of proportionality in  (\ref{barxi}).
Explicitly, we take ${\bar t}=b t_Z$, where $b>1$ is order unity.
With instabilities leading to exponential growth in field
amplitudes, we expect $b$ to depend only logarithmically on $t_Q$
and the detailed properties of the system.  We note that $b$ only
occurs in ${\tilde f}$ to the power $b^{-1/4}$, and the already
weak logarithmic dependence on the quench rate and the system is
totally diluted. That is, the efficiency is almost independent of
the system.

As a demonstration we not only assume the GL theory for
equilibrium superconductors but assume the time-dependent
Landau-Ginzburg (TDLG) equation for the transition. That is, we
take
\begin{equation}
\frac{1}{\Gamma}\frac{\partial\phi_{a}}{\partial t} =
-\frac{\delta F}{\delta\phi_{a}} + \eta_{a}, \label{tdlg}
\end{equation}
for the GL quadratic energy $F$, where $\eta_{a}$ is Gaussian
thermal noise, whose strength is irrelevant in defining $f(r,t)$
in the linear regime. This is a reliable guide to the non-linear
theory  of low-$T_c$ superconductors\cite{calzetta}. Neglecting
thermal fluctuations, we find \cite{rayulti,ray} that, in the
linear regime,
\begin{equation}
g(r,t)\approx \exp (-r^2t_0/8\xi_0^2t), \label{f1}
\end{equation}
for small $r$, provided $t$ is sufficiently large. As a result,
$g(r,{\bar t})\approx \exp (-r^2/8b\xi^2_Z)$ at time ${\bar t}$,
whereby the efficiency factor ${\tilde f}$ of (\ref{DeltaN}) takes
the value ${\tilde f}\approx b^{-1/4}/6$.   For the linear
equation (\ref{tdlg}) $b$ can be estimated from the parameters of
the model \cite{rayulti}. Values of $b$ between 2 and 3 are
plausible, which gives an efficiency roughly half the canonical
value of ${\tilde f} =1/4$, but still larger than that seen in
\cite{technion2}. What we are saying is that, although we would
not defend (\ref{tdlg}) as describing the dynamics of high-$T_c$,
lower efficiencies than the standard guesses based on static
descriptions are natural. This completes our discussion of
high-$T_c$ systems.

In the remainder of this letter, we shall take a parallel look at
experiments that observe vortices at the $U(1)$ symmetry-breaking
interface of liquid crystals at an isotropic-nematic (IN)
transition \cite{digal,Ray}. For uniaxal nematic liquid crystals
the bulk order parameter in the nematic phase is the director
vector, with order parameter space $RP^2 = S^2/Z_2$. This allows
for string defects with $n=\pm1/2$ windings. The transition is
first order. However, the anchoring of the director at the
two-dimensional IN interface forces it to lie on a cone, whereby
the order parameter space is a circle $S^1$, and defects have
winding number $n=\pm 1$, corresponding to $U(1)$ breaking. It is
suggested in \cite{digal,Ray} that the transition could evolve by
the same growth of unstable modes  that we have invoked for
superconductors. At least, the whole formation takes place almost
instantaneously
 and the structure is completely well focussed, implying that
the transition is not first order. We assume that the order
parameter field is softened by fluctuations so that defects can be
identified with its zeroes. Even if there is a crossover, rather
than a true transition, our previous analysis goes through
unchanged provided it is not too broad \cite{bettencourt}, which
we assume.

In Ref.\cite{digal} the authors cool a droplet of liquid crystal
through its critical temperature and measure $N$ and $\Delta N$
through square regions at the surface, where the behaviour is as
above.  They do not attempt to confirm (\ref{xidef}) directly, but
compare the ensemble averages $\Delta N$ to $N$. A fit to the form
\begin{equation}
\Delta N =  c (N)^{\nu}
\end{equation}
 gives $\nu = 0.26\pm 0.11$,  and a value for $c$ of
 \begin{equation}
 c = 0.76\pm 0.21.
 \label{c}
 \end{equation}

We observe that, with the same choice of correlation function (\ref{f}), the
total number $N$ of vortices and antivortices passing through a
surface, area $A$ is, from (\ref{density}),
\begin{equation}
N = A/2\pi{\bar\xi}^2,
\end{equation}
{\it irrespective} of shape (the factor of $2\pi$ in the
denominator is a direct reflection of the same factor in
(\ref{density}), and does not assume a circular surface). For a
{\it square} surface (with $A=(C/4)^2$) this enables us to use
(\ref{step2}) to eliminate ${\bar\xi}$, giving
\begin{equation}
\Delta N \approx 0.77 (N)^{1/4},
 \label{NDN}
\end{equation}
independently of $\sigma$.

This value is in remarkable agreement with $c$ of
 (\ref{c}), and is our second result. It is to be contrasted with the values of $c=0.57$
 and $c=0.71$ that are obtained under the simplistic assumption of
 random phase distribution in uniform arrays of triangular or
 square domains, respectively \cite{digal}.

The value $\nu = 1/4$ of (\ref{NDN}), already anticipated in
(\ref{DeltaN}), can take no other value insofar as field ordering
is driven by early Gaussian fluctuations before non-linearity
becomes important.

\smallskip
We thank Emil Polturak, Ariel Maniv and Rajarshi Ray for helpful
discussions. This work is, in part, supported by the COSLAB
programme of the European Science Foundation.

\end{document}